\documentclass{aa}
\usepackage[usenames,dvipsnames]{xcolor}
\usepackage{natbib}
\usepackage{xspace}
\usepackage{amssymb}
\usepackage{amsmath}
\usepackage{graphicx}
\usepackage{hyperref}

\def\revised#1{{#1}}

\begin{document}

\title{Surface waves in protoplanetary disks induced by outbursts: \\ Concentric rings in scattered light}  \titlerunning{Outburst-induced rings}
\authorrunning{Schneider, Dullemond \& Bitsch}
\author{A.D.~Schneider$^{1}$, C.P.~Dullemond$^{1}$ \& B.~Bitsch$^{2}$}
\institute{
  (1) Zentrum f\"ur Astronomie, Heidelberg University, Albert Ueberle Str.~2, 69120 Heidelberg, Germany\\
  (2) Max-Planck-Institut f\"ur Astronomie, K\"onigstuhl 17, 69117 Heidelberg, Germany
} \date{\today}

\abstract{
  {\em Context:} Vertically hydrostatic protoplanetary disk models are based
  on the assumption that the main heating source, stellar irradiation,
  does not vary much with
  time. However, it is known that accreting young stars are variable sources of
  radiation. This is particularly evident for outbursting sources such as EX
  Lupi and FU Orionis stars.\\
  {\em Aim:} We  investigate how such outbursts affect the vertical
  structure of the outer regions of the protoplanetary disk,
  in particular their appearance in scattered light at optical and
  near-infrared wavelengths.\\
  {\em Methods:} We employ the 3D FARGOCA radiation-hydrodynamics code,
  in polar coordinates, to compute the time-dependent behavior
  of the axisymmetric disk structure. The temperature is computed
  self-consistently and time-dependently from the irradiation flux
  using a two-stage radiative transfer method: first the direct illumination
  is computed; then the diffuse radiation is treated with the flux-limited
  diffusion method. The outbursting inner disk region is not included
  explicitly. Instead, its luminosity is added to the stellar luminosity
  and is thus included in the irradiation of the outer disk regions.
  For time snapshots of interest we insert the density structure
  into the RADMC-3D radiative transfer code and compute the appearance of
  the disk at optical/near-infrared wavelengths, where we observe
  stellar light that is scattered off the surface of the disk.\\
  {\em Results:} We find that, depending on the amplitude of the outbursts,
  the vertical structure of the disk can become highly dynamic, featuring
  circular surface waves of considerable amplitude. These ``hills'' and
  ``valleys'' on the disk's surface show up in the scattered light images as
  bright and dark concentric rings. Initially these rings are small and
  act as standing waves, but they subsequently lead to outward propagating
  waves, like the waves produced by a stone thrown into a pond. These waves
  continue long after the actual outburst has died out.
  \\
  {\em Conclusions:} Single, periodic, or quasiperiodic outbursts of the
  innermost regions of protoplanetary disks will necessarily lead to wavy
  structures on the surface of these disks at all radii. We propose that some
  of the multi-ringed structures seen in optical/infrared images of several
  protoplanetary disks may have their origin in outbursts that  occurred
  decades or centuries ago. However, the multiple rings seen at (sub-)millimeter
  wavelengths in HL Tau and several other disks are not expected to be related to
  such waves.
}

\maketitle


\section{Introduction}
With high-contrast optical and near-infrared imaging using 8-meter class
telescopes, it has become possible to obtain a detailed view of the surface
structure of protoplanetary disks. This new view of protoplanetary disks reveals
striking patterns, including single rings \citep[e.g.,][]{2017AJ....154...33A} or
multiple concentric rings \citep[e.g.,][]{2017ApJ...837..132V}, $m=2$ spirals
\citep[e.g.,][]{2017A&A...597A..42B}, warps \citep{2015ApJ...798L..44M}, and
moving shadows \citep{2017ApJ...849..143S}. Sometimes the rings seen in these
disks have a radial width that is narrower than the vertical scale height of the
disk, for example~in the case of RXJ 1615 \citep{2016A&A...595A.114D}. Such
structures are hard to explain as equilibrium structures in the disk, and are
more suggestive of a surface wave phenomenon of some kind. The question is then,
what could produce such waves?

The self-shadowing instability \citep{1999ApJ...511..896D,
  2000A&A...361L..17D, 2008ApJ...672.1183W, 2012A&A...539A..20S} could
be one possibility. This could
either lead to inward-moving waves or to stationary-state waves, depending on
the details of the disk structure and the central star.
%

In this Letter we propose an alternative explanation. We posit that
protoplanetary disks will not be in hydrostatic equilibrium in their
intermediate to outer regions ($r\gtrsim 10\,\mathrm{au}$) if the star and the
inner disk regions ($r\lesssim 1\,\mathrm{au}$) experience accretion outbursts
of the EX Lupi or FU Orionis kind. Such outbursts would, for a period of half a
year to many tens of years, heat the outer disk regions through irradiation.
These regions will thus become overpressurized and will start to expand vertically. Once the outburst is over, the outer disk regions cool again.  However, by
this time, the disk gas is already in motion, and the return to the ``normal''
temperature comes too late, so to speak, to prevent the violent vertical
oscillation of the disk gas. Given that the Kepler frequency is different at
different radii, the vertical oscillation of the gas occurs on different timescales even though the entire disk experienced the same duration of the
outburst.


We tested this hypothesis by numerical radiation-hydrodynamic simulations followed by
diagnostic radiative transfer computations to compute the optical appearance of
the disks.

\section{Model}
We use a modified version \citep{2014A&A...564A.135B, 2014MNRAS.440..683L} of
the legacy FARGO code \citep{2000A&AS..141..165M}, which is designed to
model the hydrodynamics of protoplanetary disks. This modified version
(hereafter FARGOCA) implements the stellar irradiation by radial ray-tracing of
the stellar light followed by a flux-limited-diffusion treatment of the thermal
radiation of the disk \citep{2013A&A...549A.124B}. It is implemented on a  3D
grid in spherical coordinates $(r,\theta,\phi)$. In this Letter we assume
the disk to be axially symmetric, which reduces the problem
effectively to 2D, with coordinates $(r,\theta)$. For stability
the irradiation is blocked within 7$^{\circ}$ of the midplane, mimicking
the shadow cast by the unresolved innermost disk.

Before we can start the actual simulation, we need to find as an initial
condition an unperturbed steady-state disk model. We do this by setting up an
initial guess of a density and temperature distribution, and letting it
undergo radiation-hydrodynamic relaxation. Our initial guess has a power law index $p$ for the
surface density. We then get
\begin{equation}
  \rho_{\mathrm{guess}}(r,z) = \frac{\Sigma_0}{H\sqrt{2\pi}}\left(\frac{r}{r_0}\right)^{-p}\exp\left(-\frac{z^2}{2 H^2}\right),
\end{equation}
where $z$ is the vertical distance from the midplane, approximately equal to
$z=(\pi/2-\theta)\,r$. The quantity $H=H(r)$ is an estimate of the vertical
scale height of the disk at radius $r$. We initialize the disk with a fair guess
of the starting values, assuring that this initial guess has a flaring geometry,
and then wait for the disk to reach hydrostatic equilibrium with irradiation and
heating.

We use reflective boundaries in the $\theta$-direction and in the radial direction at
the inner edge of the disk. The radial outer boundary remains open. For
numerical stability we introduce a floor density of
$\rho_{\mathrm{floor}}=2\times 10^{-17}\,\mathrm{g}/\mathrm{cm}^3$.

We use a grid with 500 logarithmic spaced cells in $r$-direction between 5.2 and
400 au. For $\theta$ we use 100 linearly spaced cells between $\theta=\pi/3$ and
$\theta=\pi/2$ (the midplane), i.e.,~we model only the upper half of the disk.

For our initial guess setup we chose the surface density power law index $p=1$,
and a surface
density value at the fiducial radius $r_0=5.2\,\mathrm{au}$ of
$\Sigma(5.2\,\mathrm{au})=160\,\mathrm{g}/\mathrm{cm}^2$. For the star we chose
$M_{*}=M_\odot$, $R_{*}=1.5\,R_\odot$, and $T_{*}=4370\,\mathrm{K}$. The opacity
model used for the radiation-hydrodynamics is that from
\citet{1994ApJ...427..987B}.

After relaxing the model to a quiescent state, requiring $\approx 20000$ years of
model time, we activated the outburst. Since the outburst occurs in the disk well
within the inner radius of our model, we treated the outburst simply as an
increase in the stellar luminosity by a given factor, lasting for a given time,
after which the star returns to its original luminosity. For numerical stability
we implemented a smoothed increase and decrease in luminosity. We then followed the
evolution for over 1000 years after the outburst had ceased. 

To compute scattered-light images, we post-processed our hydrodynamical simulation
using the code package RADMC-3D \citep{2012ascl.soft02015D}. For a given
distribution of dust, RADMC-3D calculates the scattered light image that 
would be observed using an instrument such as VLT/SPHERE. We calculated the
dust opacities with a Mie code using the \citet{2003ApJ...598.1017D} optical
constants, assuming a Gaussian distribution of dust grains with a mean size
of $10^{-5}\mathrm{cm}$ and a width of $5\%$.

Because the opacities of \citet{2003ApJ...598.1017D} and
\citet{1994ApJ...427..987B} are not fully compatible with each other, we
increased the canonical dust-to-gas ratio for the RADMC-3D model by a factor of
$1.42$ (leading to a dust-to-gas ratio of $0.0142$), such that the
Rosseland mean opacity computed from the \citet{2003ApJ...598.1017D} matches
that of \citet{1994ApJ...427..987B} for temperatures below about 150 K. In other
words, instead of adding water ice to the opacity,  for simplicity we adjusted the silicate
abundance. Likewise, we adjusted the ``stellar opacity'' (the
opacity of the dust grains at stellar wavelengths) used in FARGOCA to be
consistent with the \citet{2003ApJ...598.1017D} opacity, leading to a gas+dust
absorption opacity for stellar light at $\lambda\simeq 0.66\,\mu\mathrm{m}$ of
$\kappa=25.8\,\mathrm{cm}^2/\mathrm{g}$.

\section{Results}
\begin{figure}
  \centerline{\includegraphics[width=0.45\textwidth]{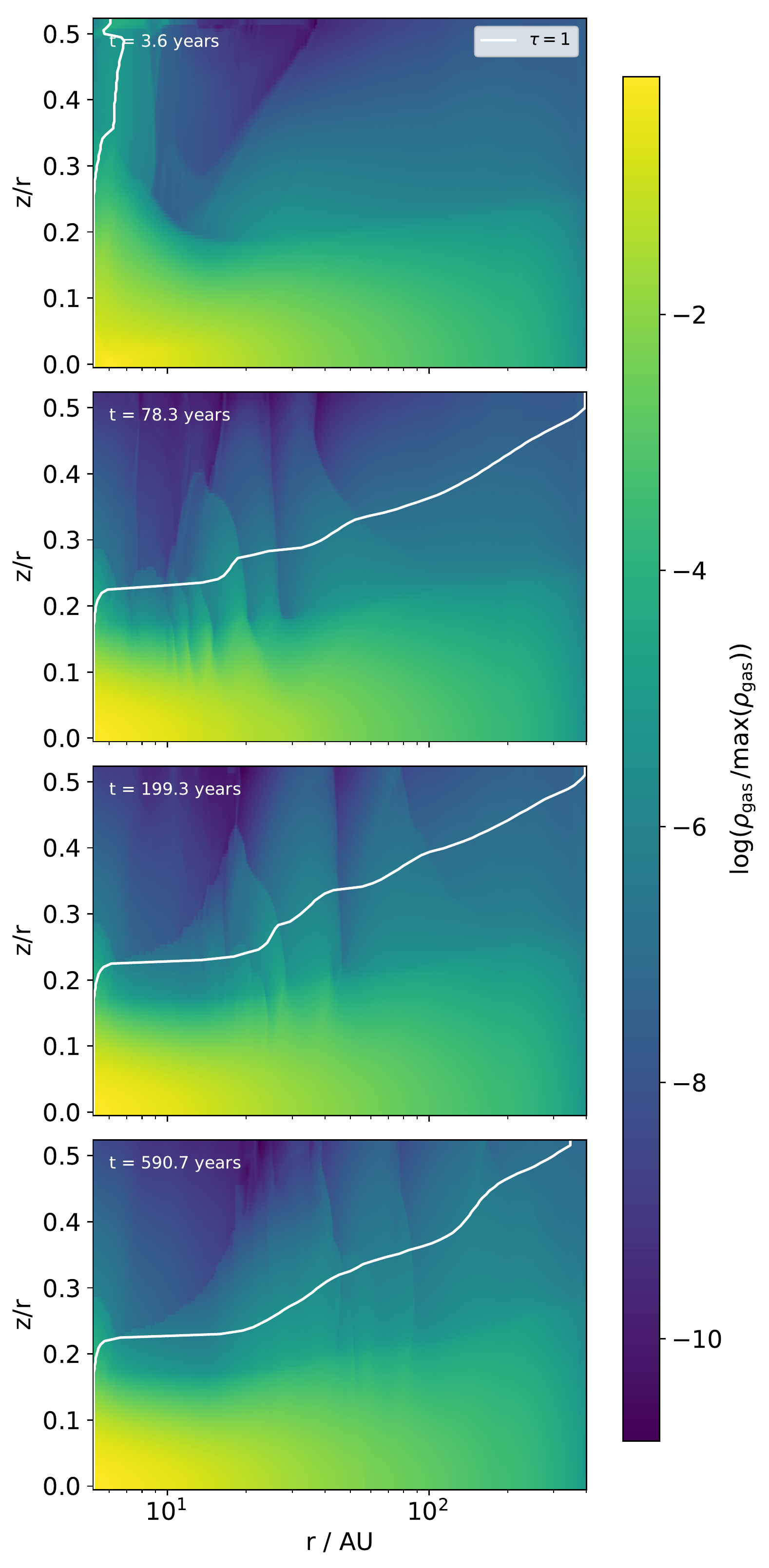}}
  \caption{\label{fig-density}Density structure of the disk at different
    epochs after the start of the outburst. The horizontal axis is the radial
    coordinate in $\mathrm{au}$. The vertical axis is $\pi/2-\theta\simeq z/r$.
    Stellar photons move horizontally in this plot. The color
    scale is logarithmic, normalized to the maximum density of the first
    image. The white line is the radial $\tau=1$ line at
    $\lambda=0.66\,\mu\mathrm{m}$ with scattering included. The albedo at this
    wavelength is 0.8. The radiation-hydrodynamics simulations,
    however, only use the absorption opacity.
  }
\end{figure}

We show the results for an outburst with a duration of 1.89 years, during which
the luminosity increases by a factor of \revised{30}. The resulting radial-vertical
density structure of the disk at different epochs after the start of the
outburst is shown in Fig.~\ref{fig-density}. It can be seen that the outburst,
leading to heating of the disk, causes strong vertical expansion of the surface
layers of the disk in a wavy manner. Once in motion, the expansion continues
even after the end of the outburst. The wavy character of the expansion is, at
least in part, due to the fact that at any radius $r$ the disk reacts roughly on
the local Kepler timescale. While the outer disk regions are still vertically
expanding, the inner regions have already collapsed and may already be bouncing
back up. The vertical expansion leads to a vertical displacement of the
$\tau=1$ surface of the short-wavelength radiation from the star. The wavy
radial behavior of the vertical expansion/collapse/bounce-back thus leads to
``mountains'' and ``valleys'' on the disk surface.

Soon after the disk has been set in vertical motion, the waves also propagate
outward. This wave is a combination of pressure wave (sound wave) and surface
wave (gravity wave). In particular in the surface layers these waves are
shocks. Their physical character is the same as the spiral waves induced by
companions or planets, but they are radially outward moving, and manifest
themselves as one or multiple concentric rings.


\revised{As can be seen in this figure, after about 200 years the inner regions of the
disk (inward of 10 au) have calmed down, but the wave is still propagating in
the regions beyond that. After 590 years the surface layers of the outer disk
regions (100 -- 400 $\mathrm{au}$) are still vertically expanding due to the
outburst heating event and the outward propagating wave has reached about 90
$\mathrm{au}$.}


\begin{figure}
  \centerline{\includegraphics[width=0.4\textwidth]{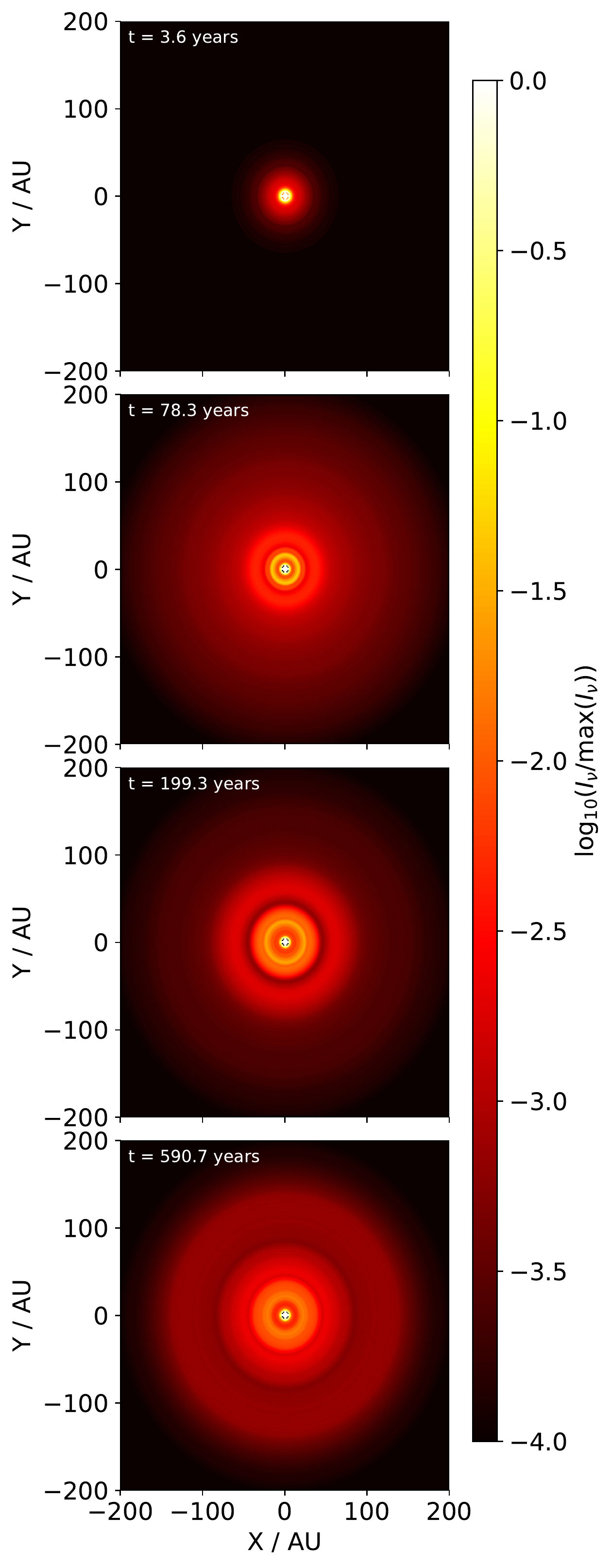}}
  \caption{\label{fig-scatlight-images}Scattered light images of the disk
    seen face-on at the same epochs as in Fig.~\ref{fig-density}. The color
    scale is logarithmic, normalized to the maximum of the first image.}
\end{figure}

The scattered light images computed for the same time snapshots as in
Fig.~\ref{fig-density} are shown in Fig.~\ref{fig-scatlight-images}. They show
that the ringlike mountains and valleys  have sunny sides and shadowed
sides, visible as bright and dark rings in the images. \revised{The contrast between
peaks and valleys in the image are on the order of a factor of two to four.} For
comparison, the contrast of the rings seen in the scattered light image of TW
Hydra \citep{2017ApJ...837..132V} is \revised{on the order of a factor of two}, and for
RXJ 1615 \citep{2016A&A...595A.114D} it is, \revised{depending on the ring/valley pair, on the order of tens of percent to a factor of 5 to 10.}
The order of magnitude of the ring contrast thus appears to match the
observations of these sources.

\revised{It should be noted, however, that the outburst amplitude of the
  fiducial model shown in this Letter is a bit large for an EX Lupi-like
  outburst. In EX Lupi itself, the bolometric luminosity amplification factor
  during the 2008 outburst was on the order of 10, even though in the V band it
  was up to a factor of 100 \citep{2009Natur.459..224A}. We also made a model
  calculation with luminosity factors of 5 for which we found contrast ratios
  on the order of 1.3 up to $\sim$2. FU Orionis outbursts can be very bright,
  but they also last longer. Our models also find circular waves in those
  cases.}

The rings in our model images move outward with time. The outward speed of the
rings is about 1 km/s amounting to about 0.25 $\mathrm{au}/\mathrm{year}$ for
the current setup. For broad rings, such as those seen in TW Hydra, it may take
a few years between consecutive images to be able to discern the outward
motion. For narrow rings, such as the rings seen in RXJ 1615, the time baseline
may be smaller.

\section{Discussion}
A strong prediction of this model is that the rings seen in scattered light
should move radially outward, meaning that over a time baseline of several
years a discernable change in the radius of the rings should be detectable. If
the rings observed in a specific source can be shown to be static on a timescale on the order of half the wavelength (crest-to-crest radial distance)
divided by the sound speed, then the model is disproven for that source.


Measurement of the radial velocities caused by the outward propagating waves
using CO molecular lines with ALMA would provide strong evidence for the wave
scenario, but the amplitude of these radial velocity waves depends strongly on
the strength and duration of the outburst, and it requires a more
detailed study to determine which radial velocities are  expected for a given
amplitude of the bright and dark rings in the scattered light images.

If our scenario is correct for a particular source, then the radial velocity
pattern, as measured in different CO isotopologues, may also offer the
opportunity to do tomography of the disk in a way similar to the recent work by
\citet{2018MNRAS.474L..32J}. It would also allow us to compute the time in the
past, the duration, and the strength of the putative outburst.

\revised{The present models are still rather simplified. Given that the
  phenomenon is driven by irradiation, a better treatment of the dust opacities,
  possibly with dust settling included, will be necessary in future work. The
  effect of multiple smaller outbursts should be investigated, and in general a
  detailed parameter study with comparison to sources on a case-by-case
  basis should be undertaken.}

\section{Conclusions}
We have shown that the sudden luminosity increase of the central regions of a
protoplanetary disk will cause long-lasting surface waves in the disk's outer
regions. The sudden irradiative heating by the luminosity burst drives the disk,
or at least its surface layers, out of vertical hydrostatic equilibrium. The
ring-shaped mountains and valleys that are formed are seen in scattered light as
alternating bright and dark rings. We speculate that these multi-ringed patterns
in scattered light may explain some of the multi-ringed scattered light patterns
seen in several nearby protoplanetary disks, such as TW Hydra
\citep{2017ApJ...837..132V} and RXJ 1615 \citep{2016A&A...595A.114D}.

Clearly, strong enough outbursts of EX Lupi or FU Orionis type must cause
circular waves in the disk in the way described in this Letter. Whether the
rings we see in scattered light in several nearby disks are caused by such
outbursts remains to be determined, but if they are not due to this mechanism,
then it would be worthwhile to search for these patterns in disks that are known
to have experienced outbursts in the past.

\begin{acknowledgements}
  C.P.D. thanks Antonella Natta for discussions that motivated this work.  We
  acknowledge support from the High Performance and Cloud Computing Group at the
  Zentrum f\"ur Datenverarbeitung of the University of T\"ubingen, the state of
  Baden-W\"urttemberg through bwHPC, and the German Research Foundation (DFG)
  through grant no. INST 37/935-1 FUGG. We also acknowledge support from DFG grant
  DU 414/23-1, which is part of the Forschergruppe FOR 2634 ``Transitional
  Disks''. B.B.\ thanks the European Research Council (ERC Starting Grant
  757448-PAMDORA) for their financial support. We thank the anonymous referee
  for the useful remarks that helped improve the manuscript.
\end{acknowledgements}

\begingroup
\bibliographystyle{aa}
\bibliography{ms}
\endgroup

\end{document}